\documentclass[twocolumn,groupedaddress,superscriptaddress]{aastex631}
\usepackage{amsmath}
\usepackage{booktabs}
\usepackage{array}

\begin{document}

\title{Efficient Stochastic Template Bank using Inner Product Inequalities}

\correspondingauthor{Keisi Kacanja}
\email{kkacanja@syr.edu}

\author[0009-0004-9167-7769]{Keisi Kacanja}
\affiliation{Department of Physics, Syracuse University, Syracuse, NY 13244, USA}

\author[0000-0002-1850-4587]{Alexander H. Nitz}
\affiliation{Department of Physics, Syracuse University, Syracuse, NY 13244, USA}

\author[0000-0002-9188-5435]{Shichao Wu}
\affiliation{Max-Planck-Institut für Gravitationsphysik (Albert-Einstein-Institut), D-30167 Hannover, Germany and Leibniz Universität Hannover, D-30167 Hannover, Germany}

\author[0000-0003-4075-4539]{Marco Cusinato}
\affiliation{Departament d'Astronomia i Astrofísica, Universitat de Valencia, Edifici d’Investigaciò Jeroni Munyoz, C/ Dr. Moliner, 50, E-46100 Burjassot (València), Spain}

\author[0000-0002-5077-8916]{Rahul Dhurkunde}
\affiliation{Max-Planck-Institut für Gravitationsphysik (Albert-Einstein-Institut), D-30167 Hannover, Germany and Leibniz Universität Hannover, D-30167 Hannover, Germany}

\author[0000-0002-5304-9372]{Ian Harry}
\affiliation{University of Portsmouth, Institute of Cosmology and Gravitation, Portsmouth PO1 3FX, United Kingdom}

\author[0000-0001-5078-9044]{Tito Dal Canton}
\affiliation{Université Paris-Saclay, CNRS/IN2P3, IJCLab, 91405 Orsay, France}

\author[0000-0002-7537-3210]{Francesco Pannarale}
\affiliation{Dipartimento di Fisica, Università di Roma ``Sapienza'', Piazzale A. Moro 5, I-00185, Roma, Italy}
\affiliation{INFN Sezione di Roma, Piazzale A. Moro 5, I-00185, Roma, Italy}

\begin{abstract}
Gravitational wave searches are crucial for studying compact sources like neutron stars and black holes. Many sensitive modeled searches use matched filtering to compare gravitational strain data to a set of waveform models known as template banks. We introduce a new stochastic placement method for constructing template banks, offering efficiency and flexibility to handle arbitrary parameter spaces, including orbital eccentricity, tidal deformability, and other extrinsic parameters. This method can be computationally limited by the ability to compare proposal templates with the accepted templates in the bank. To alleviate this computational load, we introduce the use of inner product inequalities to reduce the number of required comparisons. We also introduce a novel application of Gaussian Kernel Density Estimation to enhance waveform coverage in sparser regions. Our approach has been employed to search for eccentric binary neutron stars, low-mass neutron stars, primordial black holes, supermassive black hole binaries. We demonstrate that our method produces self-consistent banks that recover the required minimum fraction of signals. For common parameter spaces, our method shows comparable computational performance and similar template bank sizes to geometric placement methods and stochastic methods, while easily extending to higher-dimensional problems. The time to run a search exceeds the time to generate the bank by a factor of $\mathcal{O}(10^5)$ for dedicated template banks, such as geometric, mass-only stochastic, and aligned spin cases, $\mathcal{O}(10^4)$ for eccentric and $\mathcal{O}(10^3)$ for the tidal deformable bank. With the advent of efficient template bank generation, the primary area for improvement is developing more efficient search methodologies.

\end{abstract}
\section{Introduction} \label{sec:intro}

Gravitational waves (GWs) are distortions in spacetime. One mechanism for generating GWs is accelerating two massive compact objects such as neutron stars and black holes 
\citep{einstein1916,einstein1918}.  When these compact objects form a binary system, their mutual gravitational attraction causes them to spiral towards each other, producing gravitational waves that travel outward at the speed of light. We observe these GWs with ground-based detectors such as the Advanced Laser Interferometer Gravitational-wave Observatory (LIGO) and Advanced Virgo \citep{2015CQGra..32b4001A,2015CQGra..32g4001L}. Since the commencement of the advanced detector era in 2015, Advanced LIGO and Virgo have detected $\mathcal{O}(100)$ gravitational wave sources \citep{2019PhRvX...9c1040A,PhysRevD.109.022001,PhysRevX.13.041039,2023ApJ...946...59N,Mehta:2023zlk,Olsen:2022pin}

The initial identification of gravitational-wave sources involves conducting searches \citep{2016CQGra..33u5004U,2021SoftX..1400680C,2021CQGra..38i5004A,Messick:2016aqy,2020PhRvD.101j4055P,2020arXiv201106787C}. Model based searches typically employ matched filtering which compares the detector data with modeled GW signals \citep{PhysRevD.85.122006,PhysRevD.87.024033}. Since the properties of a potential source are unknown, a discrete set of template waveforms, known as template bank, is used. 

Template banks are carefully constructed to ensure minimal loss in the signal-to-noise ratio (SNR) \citep{2012PhRvD..85l2006A,PhysRevD.44.3819,PhysRevD.53.6749,PhysRevD.60.022002,1996PhRvD..53.3033B,2009PhRvD..80j4014H,2012PhRvD..86h4017B,2014PhRvD..90h2004D,PhysRevD.99.024048,PhysRevD.108.042003}. An ideal template bank contains at least one sufficiently similar template waveform to identify any potential signal within a designated parameter space. A valid template bank algorithm strategically fills the parameter space with templates to limit loss in SNR, and minimizes redundant templates. Additionally, it should adapt to the specific characteristics of the gravitational-wave source population studied, allowing the ability to analyze data for a diverse range of astrophysical scenarios.

Various techniques exist for constructing a template bank. The two main classes of methods are geometric \citep{PhysRevD.86.084017,2006CQGra..23.5477B,2024PhRvD.109d2005S}, stochastic placement techniques \citep{2009PhRvD..80j4014H,2014PhRvD..89h4041A,2010PhRvD..81b4004M,Babak:2008rb}. There also exist hybrid methods that utilizes both approaches. \citep{2017arXiv171108743R,Roy:2017qgg}.  Geometric methods typically place template waveforms using a lattice technique; this generally requires a known metric, which directly quantifies the similarity of signals. Afterwards, the templates are converted into a physical parameter space. This approach guarantees that every template captures a predefined minimum percentage of the signal \citep{PhysRevD.53.6749,2007CQGra..24S.481P}. One challenge is that the metric may only be flat in the non-physical space. This may result in over-placing templates near and outside the physical space boundaries. These templates need to be included since they may be better at recovering signals near the boundary than the templates within the physical space \citep{2007PhRvD..76j2004C,2022PhRvD.106l2001C,2012PhRvD..86h4017B}. Consequently, since a search's computational cost is proportional to the number of templates produced from any method, including these templates can lead to an excess in the template count, making the search more expensive to perform. This approach is not optimal for complicated boundaries, unknown metrics, and for more intricate detection studies requiring non-trivial parameter spaces.

An alternative approach is a stochastic placement technique. This method can work without an explicit metric and does not attempt to construct a lattice to place templates. Instead, templates are placed directly in the physical space and iteratively added to the bank until the parameter space is sufficiently well covered. There are several approaches for stochastic template proposals and how templates are evaluated for inclusion in a bank. Stochastic methods typically propose templates either randomly or according to a probability distribution \citep{2014PhRvD..89b4003P,2017PhRvD..95f4056I}. Various template acceptance strategies exist, such as rejecting templates that retrieve more than a predetermined percentage of the SNR. Alternatively, some methods gauge the density of templates; if the volume of templates surpasses a specified threshold, the template bank is deemed adequately covered \citep{PhysRevD.80.104014}. Unlike geometric methods, stochastic methods provide less stringent guarantees on the completeness of the template bank, sometimes resulting in banks with small holes with lower accuracy of SNR reconstruction. However, stochastic methods in general offer a higher degree of flexibility than geometric methods, and allows the ability to perform searches for many variety of circumstances such as eccentric or deformable sources.

In a broader scientific context, the construction of template banks can be likened to the principles of experimental design,  where the goal is to efficiently sample a high-dimensional parameter space, similar to methods used in statistics, computer science, and manufacturing for uncertainty quantification and optimization \citep{10.1214/ss/1177012413}. Each experiment in these fields corresponds to a specific set of input parameters, and the objective is often to understand the variation of a Quantity of Interest (QoI) based on these parameters, ensuring adequate space-filling to capture the underlying behavior of the system \citep{10.1214/ss/1177012413}. Similarly, template banks for gravitational wave detection aim to cover the parameter space as fully as possible to maximize detection efficiency and minimize missed signals.

In this paper, we present a new stochastic placement method. We show this method produces self-consistent banks that recover the SNR of any potential signal within a target seqarch space, with losses below a chosen threshold. We show that this method has already been used to conduct various searches such as (low-mass and eccentric) neutron stars, (supermassive and primordial) black holes, and neutron-star–black-hole binaries that produce gamma-ray bursts. Finally, we show how the number of parameters in a bank scales the time to complete a search by generating different banks with varying intrinsic parameters such as spin, eccentricity, and or tidal deformability.

\section{Evaluation of Template Bank Coverage} \label{sec:performance} 

We assess the effectiveness of a template bank by determining the fraction of SNR its best matching template can recover for a given source, known as the fitting factor (FF) \citep{PhysRevD.52.605}.
To quantify the similarity between template waveforms, and the ability of one GW waveform to recover the SNR of another, we define the overlap between waveforms to be
\begin{equation}\label{eq:overlap}
\mathcal{O}\left( h_0, h_1 \right) = \frac{\left( h_0 | h_1 \right)}{\sqrt{\left( h_1 | h_1 \right) \left( h_0 | h_0 \right)}}
\end{equation}
The noise-weighted inner product $(h_0|h_1)$ is the fraction of signal power extracted from a modeled signal $h_1$ with the waveform model $h_0$, defined as
\begin{equation}
\left( h_0 | h_1 \right) = 4 \, \mathrm{Re} \int_{0}^{\infty} \frac{\tilde{h}_0(f) \tilde{h}_1^*(f)}{S_n(f)} \, df
\end{equation}
Where $S_n(f)$  is noise power spectral density and $f$ is frequency.
Since an overall phase and the absolute time of arrival are typically nuisance parameters in a search, we define the phase and time maximized overlap to be the match $\mathcal{M}$. This formalism is for a non-precessing signal. For precessing signals,  refer to \citet{2016APS..APRK14005H}. \\
\begin{equation}
\mathcal{M} \left( h_0, h_1 \right) = \max_{\phi_c, t_c} (\mathcal{O}\left( h_0, h_1 \right (\phi_c, t_c))
\end{equation}
The match measures how well the observed waveform correlates with the expected waveform for a given the detector sensitivity. The maximum match for a potential signal $h_i$ with all the templates in a bank $h_{tb}$ is the fitting factor.  The same waveform approximant is used for both templates when testing the coverage of the bank.

\begin{equation}
FF(h_i) = \max_{h \in \{h_{tb}\} } \mathcal{M}(h_i, h)
\end{equation}
The FF represents the fraction of the SNR that is recovered by the template bank for a given signal. For instance, consider a template bank constructed with a minimal match of 0.95. This bank is expected to recover signals with FFs of 0.95, such that up to 5\% of a signal's SNR or a maximum of 14\% of the total number of signals may be sacrificed. FFs less than 0.95 indicate regions within the bank where templates cannot fully recover the SNR of a reference signal, suggesting potential gaps in coverage. FFs at 0.95 and above signify that the bank is adequately populated with templates, capable of recovering at least 95\% of the SNR.

\section{Methods: Stochastic Template Bank Algorithm} \label{sec:alg}
\begin{figure*}
    \hspace*{-1.5cm}
    \centering
    \includegraphics[height = 16cm, width = 21cm]{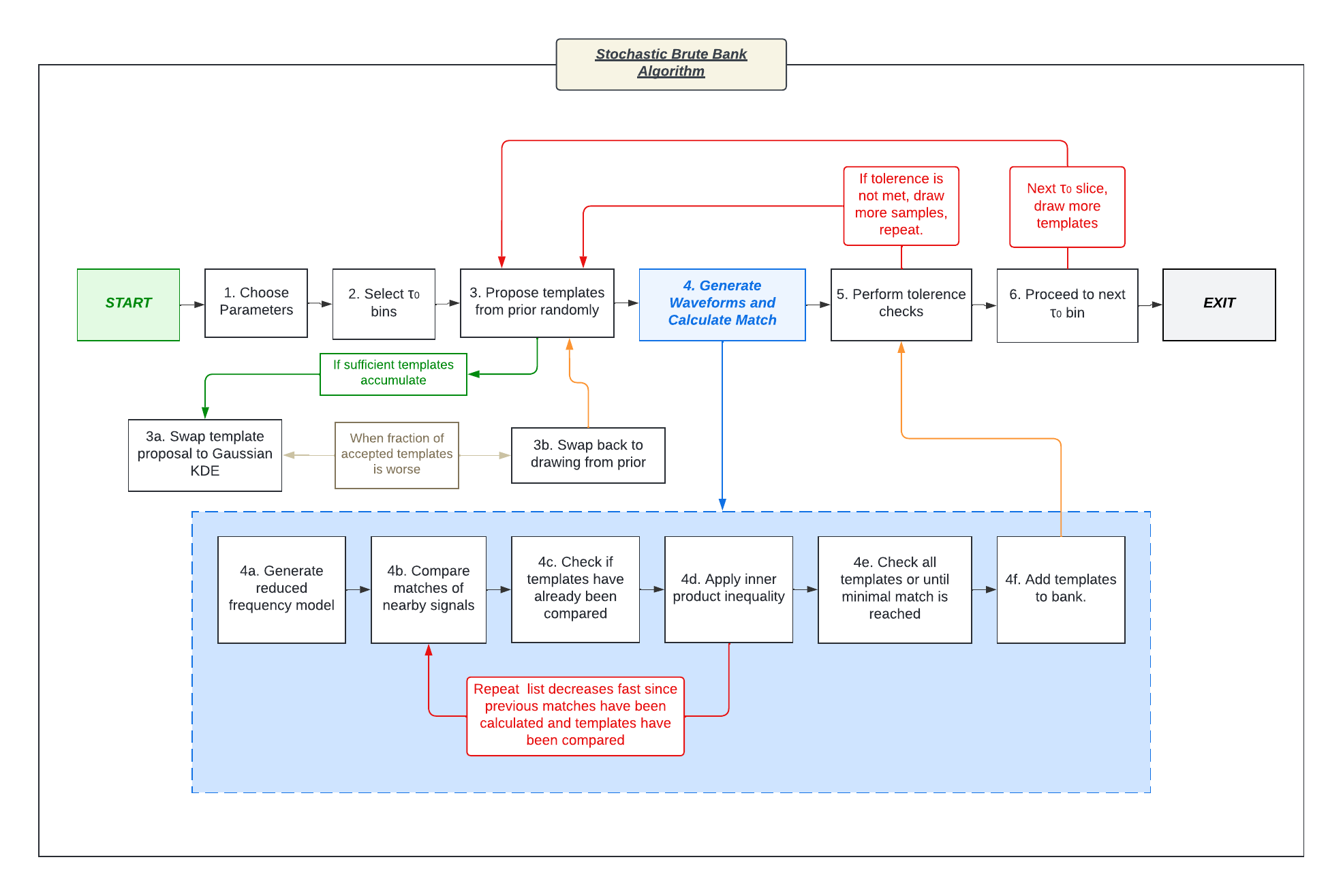}
    \caption{Stochastic Bank Algorithm Diagram. Points are randomly sampled from the prior distribution and within distinct bins defined by chirp time $\tau_0$. Waveforms are then generated with a reduced frequency model. When a template is proposed, we assess if including the proposal in the template bank will improve the coverage of the desired parameter space. This assessment is outlined in the blue box and involves comparing the proposed template with the nearest neighboring templates in chirp time. Templates are added to the bank if match between the proposed template and the accepted templates fall below the minimal match condition. If the match exceeds the minimal match condition, the bank is deemed adequately covered in that region. Match calculations are stored to streamline the process and avoid redundant computations for new proposed templates, utilizing a triangle inequality for efficient comparisons. Upon completing comparisons and reaching the minimal match threshold, tolerance checks, outlined in red, ensure the bank's coverage adequacy. If the tolerance condition is unmet, additional samples are drawn either stochastically from the prior distribution or with a Gaussian Kernel Density Estimation (KDE) method. The methods are swapped if one technique performs better than the other depicted in the beige box. Proposals are once again compared with the accepted templates. Once the tolerance condition is satisfied, accepted templates are saved, and the process iterates for subsequent bins until the entire bank is sufficiently covered. The red arrows represent the checks and repetition in method required to fulfill either tolerance or minimal match conditions.} 
    \label{fig:alg}
\end{figure*}

Our method uses a stochastic approach to place templates directly in the physical space and enables the generation of templates that covers a wide variety of parameter spaces includes those parameterized by mass, spin, tidal deformability, and eccentricity. If templates include detector responses, this method can also cover those extrinsic parameters. A diagram of the algorithm is illustrated in Figure \ref {fig:alg}, where it is divided into two main processes: the mechanism which proposes templates (discussed in section \ref{sec:sa}), and the optimized procedure which determines whether a proposed template should be included into the bank (discussed in section \ref{sec:tf}). Overall, this method proposes templates and determines whether the inclusion of the templates sufficiently cover the parameter space to recover any potential signal and will repeat until the desired coverage is achieved. 

The bank is sufficiently covered under two termination criteria: minimal match and tolerance. Tolerance is the termination criteria threshold defined by the user. When the tolerance falls below the acceptance fraction, the fraction of accepted versus the proposed templates, the region of a template bank is considered to be completed. This criterion determines how many templates must be drawn to sufficiently cover the bank. Lower tolerance values generate banks with more templates to cover the parameter space adequately; larger tolerance values do the opposite. The minimal match is the minimum percentage of SNR that at least one template in the bank can recover any fiducial signal. We use this criterion to accept or reject proposal templates. If the match between the proposal and the accepted templates is less than the minimal match, that proposed template will be added to the bank. If the template has a match greater than the minimal match condition, sufficient templates exist in the bank to recover any given signal in that region.

\subsection{Template Sampling Algorithm}\label{sec:sa}
In this section, we detail the template sampling method, the first stage of our stochastic template bank method. We first initialize the method with a set of user-defined parameters such as properties of the source, the lower frequency cutoff, and the PSD model. The sample templates are drawn within the bounds of chirp time. Chirp time ($\tau_0$), known as the zeroth order signal duration in time, defines the regions of template placement \citep{2006CQGra..23.5477B}. We utilize $\tau_0$ bins for sampling nearby templates, as significant differences in chirp time between signals would lead to higher mismatches. Moreover, stochastic methods struggle to parallelize the physical space sampling over multiple cores since splitting up the parameter space can be challenging \citep{PhysRevD.106.122001}. Utilizing $\tau_0$ boundaries facilitates faster template bank generation by parallelizing different $\tau_0$ bins along different cores. $\tau_0$ is defined as follows

\begin{equation}
\tau_0 = \frac{a_0(f_{\text{lower}})}{{\mathcal{M}_{\text{chirp}}}}^{5/3}
\end{equation}
where $a_0$ is 
\begin{equation}
    a_0(f_{\text{lower}}) = \frac{5}{{256 (\pi f_{\text{lower}})^{8/3}}}
\end{equation}

and $\mathcal{M}_{\text{chirp}}$ is defined as

\begin{equation}
    \mathcal{M}_{\text{chirp}} = \frac{(m_1m_2)^\frac{3}{5}}{(m_1+m_2)^\frac{1}{5}}
\end{equation}
where $m_1$ and $m_2$ is the mass of primary and secondary respectively.

\begin{figure}
    \centering
    \vspace*{-0.5cm}
    \includegraphics[height =20.2cm, width =7.8cm]{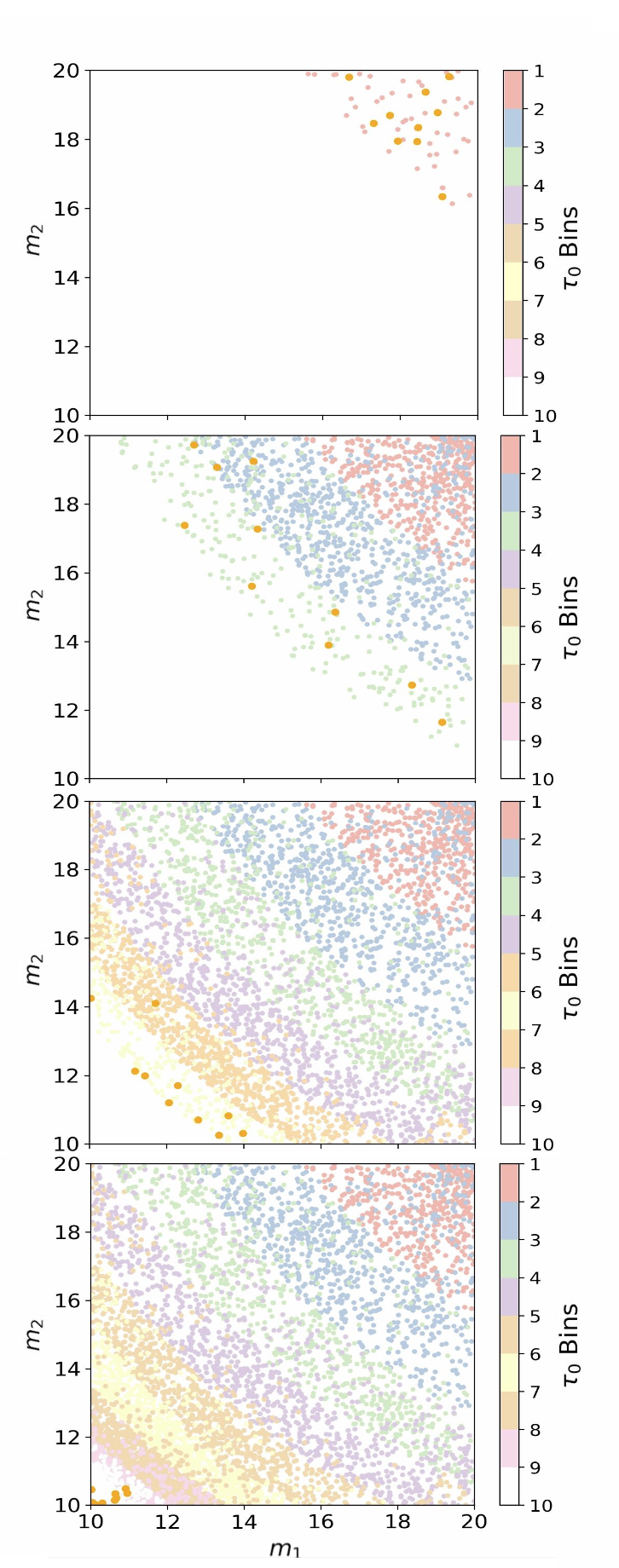}
    \caption{The evolution of the template bank as proposed templates are added within bounded $\tau_0$ bins. Once the first $\tau_0$ region is sufficiently filled, the algorithm proceeds to the next region, which is overlapped with the previous bin by 50\%. Points are proposed until the whole bank is sufficiently filled. The orange points correspond to the proposed points. The small colored points correspond with the accepted templates. The color bar represents the $\tau_0$ bins explored by the method. Bank generation can be parallelized by generating it across different $\tau_0$ bins.}
    \label{fig:ani}
\end{figure}

To define the $\tau_0$ boundaries, we calculate the chirp time defined above using a default lower chirp time frequency of 15Hz for the minimum and maximum mass range corresponding to the start and end $\tau_0$. The start $\tau_0$ will correspond to the highest masses in the space and the end will correspond to the lowest mass pair. The crawl parameter determines the width of each bin. Once the $\tau_0$ boundaries are defined, we draw the first set of points from the prior distribution of the parameters within the first $\tau_0$ bin. The sampled templates are either accepted or rejected into the bank (elaborated in section \ref{sec:tf}). 

The objective of an effective stochastic proposal is to ensure comprehensive coverage of the parameter space, minimizing gaps and undercoverage. An ideal stochastic proposal would be one proportional to the final template density, which is uniform in metric space. In general, one may not know this distribution a priori. To approximate the ideal distribution, and minimize the chance of gaps, holes, or undercoverage, we utilize a Gaussian Kernel Density Estimate (KDE). We construct the KDE from the most recently accepted template parameters after a sufficiently large population is collected using a chosen prior proposal distribution. The KDE distribution updates as new points are accepted and can then evolve towards the ideal distribution. However, the KDE alone cannot fully resolve issues arising from complex boundaries within the parameter space. Hence, the algorithm switches between drawing uniformly from the prior proposal and the KDE-based proposal based on which is more efficient at the time. Once the first $\tau_0$ bin has been sufficiently filled with templates and the tolerance criteria is achieved, we begin sampling in the next $\tau_0$ bin overlapping 50\% with the previous strip. Figure \ref{fig:ani} shows how the $\tau_0$ strips overlap with each bin as the sampler proceeds to cover the entire parameter space.

To parallelize the bank generation, we divide the parameter space into sub-banks, each of which can be processed independently. To achieve efficient parallelization, each sub-bank is run over a set of independent $\tau_0$ bins, chosen so that the match between templates across boundaries of adjacent sub-banks remains below the minimal match threshold. Because templates are compared across $\tau_0$ bins within the generation of each sub-bank, over-coverage only occurs in the bounding $\tau_0$ bins as they are not compared to templates across shared boundaries with other sub-banks. This means that if each sub-bank has 50 $\tau_0$ bins, then the over-coverage is limited to less than $\sim 2\%$. For most analyses, we consider this excess to be negligible.

\subsection{Optimizing Template Proposal Acceptance}\label{sec:tf}

Once a template is proposed, we must determine if it should be added to the template bank to enhance its coverage of the target parameter space. This is done by determining the FF of our proposed template against the current set of templates in the bank to assess the similarity. Templates with FFs less than the minimal match condition are not sufficiently similar and need to be added to the bank. Templates exceeding the minimal match condition are similar and do not need to be included in the bank.
To start, we generate a reduced frequency waveform model, inspired by the methodology detailed in \citet{2014PhRvD..89h4041A,PhysRevD.93.124007}. This model significantly reduces the computational load required for generating waveforms for the match calculations. PSD fluctuations that typically occur during an observing run have been found to have minimal impact on template placement and coverage; standard searches use a template bank generated on the expected or average noise curve of an observing run ~\citep{2016CQGra..33u5004U}. Next, we evaluate the matches between the proposed waveform models and the closest templates in chirp time, storing these values. To rule out templates where the match is clearly above the minimal match criterion, we make use of previously stored matches. We further optimize this procedure using triangle inequality which states that the sum of any two sides of any triangle will be greater than the third side. If we compare the distances between templates, we will automatically know that the third distance between the templates will be smaller. For example, consider the mismatch (1 - $\mathcal{M}$), which represents the distance of the templates, between waveform A and waveform B, and between waveform B and waveform C. If the sum of mismatches between A and B, B and C is smaller than the predefined 1-FF threshold, we can skip the direct calculation of the match between A and C. 

\begin{equation}
    [1-FF] \geq [1 - \mathcal{M}(A,B)] + [1-\mathcal{M}(B,C)] \geq  [1-\mathcal{M}(A,C)]
\end{equation} 

We iteratively compare the matches between templates until all waveforms are checked or a template exceeding the minimal match condition is found. We then return to the sampling procedure in section \ref{sec:sa} to conduct tolerance checks. If these checks are unmet, we repeat sampling more potential templates until the $\tau_0$ bin is sufficiently covered. If the checks are met, proceed to the next bin or finish executing if the final bin has been filled.

\section{Bank Verification} \label{sec:ff}

\begin{figure}
    \centering
    \hspace*{-0.5cm}
    \vspace*{-0.5cm}
    \includegraphics[height = 7.3cm, width =9.6cm]{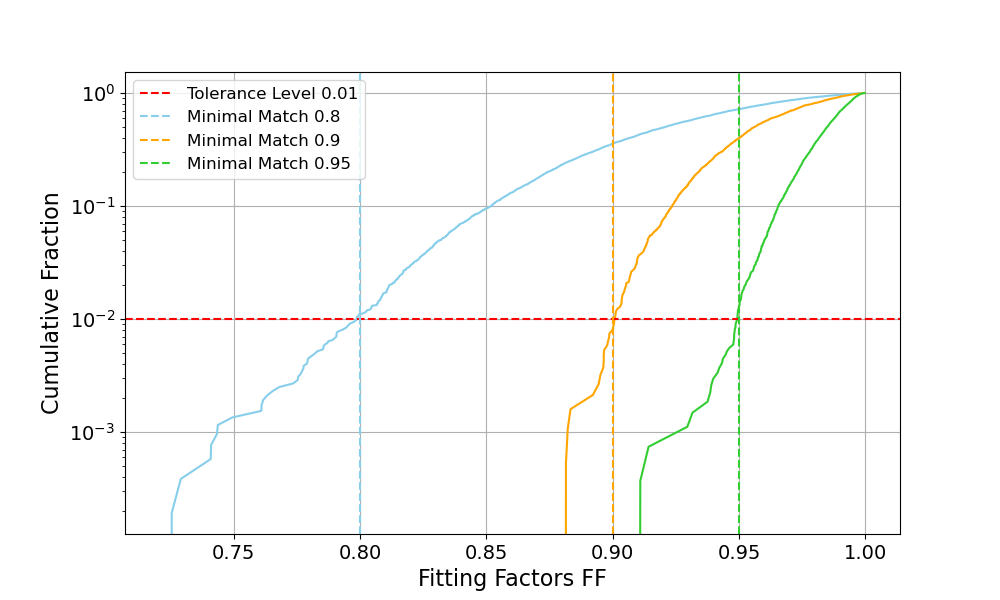}
    \hspace*{-0.5cm} 
    \includegraphics[height = 7.3cm, width =9.6cm]{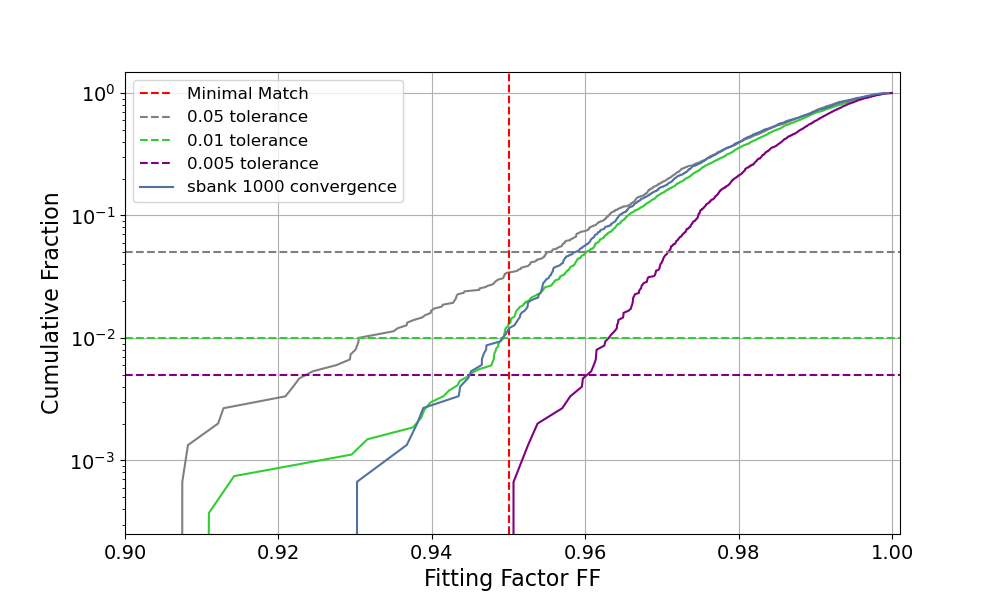}
    \caption{Cumulative distribution of fitting factors (FF)  achieved by the template banks with varying minimal match (top) and varying tolerance (bottom). The minimal matches are fixed to be 0.8, 0.9 and 0.95 in the colored vertical lines and a fixed tolerance of 0.01 in the red horizontal. The tolerances for the bottom plot vary from 0.05,0.01, and 0.005 in colored vertical dashed lines and a fixed minimal match of 0.95 in a red horizontal line. All plots achieve tolerance at the minimal match condition or before minimal match. We compare how our method performs with an existing stochastic method \citep{2014PhRvD..89h4041A,PhysRevD.93.124007} by choosing the same parameter space, minimal match of 0.95 and a convergence criteria of 1000. This bank is comparable to our method generated with a tolerance of 0.01.}
    \label{fig:ff}
\end{figure}

To demonstrate that our method produces valid template banks, we choose a fiducial parameter space and use our method to generate template banks. We perform the FF calculations described in section \ref{sec:performance} and verify the stochastic method generates self-consistent banks. We generated five separate template banks of $\mathcal{O}(10^4)$ points with fixed mass and spin range for all the banks. Masses are fixed from 2 to 10 $M_{\odot}$. Spins are fixed to be between -0.2 and 0.2. We also fixed the approximant to be IMRPhenomD, the lower frequency for the waveform to be 20Hz, and the PSD to be the Advanced LIGO final design sensitivity \citep{2020PhRvD.102f2003B,2016PhRvD..93d4006H,PhysRevD.93.044007,Husa:2015iqa,Khan:2015jqa}. To showcase the effects of tolerance choices on the bank, we generate three banks with varying tolerances, 0.05, 0.01, and 0.005, for a fixed minimal match of 0.95. We also showcase the effects of minimal match by generating the last two banks with varying minimal matches of 0.8,0.9, and 0.95 reused from the previous bank, with a fixed tolerance of 0.01. We compare our method to another stochastic method \citep{2014PhRvD..89h4041A,PhysRevD.93.124007} by generating a bank with same parameter space, minimal match of 0.95 and a convergence value of 1000. From each of these banks, we calculated the match of the templates with random fiducial waveforms for the same approximant and parameter range defined above. We successfully reconstruct a template bank that agrees with the tolerance and minimal match conditions. Figure \ref {fig:ff} shows the cumulative distribution of the recovered FF values for all six banks. All five banks generated with our method, are able to recover the fiducial signals before the minimal match and tolerance is met. \\

\newpage

\section{Computational Scaling} \label{sec:search}

\begin{table*}[hbt!]
    \centering
    \caption{The computational time to generate template banks and conduct a search for different parameter spaces. The bolded parameters correspond to template banks generated using our method. The first three banks share the same mass-only parameter space but differ in the methods used: the geometric bank \citep{2012PhRvD..86h4017B}, an existing stochastic method \citep{2014PhRvD..89h4041A,PhysRevD.93.124007}, and our method respectively. Our method produces comparable sized banks to the other methods in similar times. However, unlike the geometric and existing stochastic method, our method can generate banks with higher dimensional spaces that include eccentricity, deformability and so on. The last three rows use our method to construct template banks with spin, eccentricity and tidal deformability parameters. For searches, we assume one core processes 5000 templates in real time and estimate the time required to analyze one year of data from three detectors. The time it takes to complete a search is a factor of $\mathbf{O}(10^5)$ more than the time to complete the bank generation for the non-spinning and only spin aligned case, $\mathbf{O}(10^4)$ for the eccentric bank, and $\mathbf{O}(10^3)$ for the tidal deformable case. The non-spinning geometric technique generates banks faster than the stochastic methods for mass only banks. However, the time ratio is the largest for geometric methods, implying that currently template bank generations are a negligible cost to running a search, and faster search methods should be investigated. For larger parameter spaces, the banks generated in around  30 to 60 core days. The fraction of search time over bank generation time was the lowest for these scenarios. Overall, the most time consuming process is the search. The cost of generating the banks is trivial to the cost of running the search and faster search methodologies should be developed.}
    \label{tab:scale}
    \begin{tabular}{p{3.5cm} *{5}{>{\centering\arraybackslash}p{2.5cm}}}
    \toprule
    Parameters & Bank Size (templates) & CPU Time To Generate Bank (core days) & CPU Time to Complete Search (core days) & Search over Bank Time Ratio & Waveform Approximant \\
    \midrule
    $m_{1}, m_{2}$ (geometric) & $5.81\times 10^4$ & 0.014 & $1.27\times 10^4$ & $9.09 \times 10^5$ & Does not apply \\
    $m_{1}, m_{2}$ (sbank) & $5.84 \times 10^4 $ & 0.099 & $1.28 \times 10^4$ & $1.29\times 10^5$ & TaylorF2 \\
    $\boldsymbol{m_1}, \boldsymbol{m_2}$ & $5.29\times 10^4$ & 0.033 & $1.16\times 10^4$ & $3.52\times 10^5$ & TaylorF2 \\
    \midrule
    $\boldsymbol{m_1}, \boldsymbol{m_2}, \boldsymbol{\chi_1}, \boldsymbol{\chi_2}$ & $4.42\times 10^5$ & 0.202 & $9.68\times 10^5$ & $4.79\times 10^5$ & TaylorF2 \\
    $\boldsymbol{m_1}, \boldsymbol{m_2}, \boldsymbol{\chi_1}, \boldsymbol{\chi_2}, \boldsymbol{e}$ & $5.67\times 10^6$ & 32.23 & $1.24\times 10^6$ & $3.85\times 10^4$ & TaylorF2ecc \\
    $\boldsymbol{m_1}, \boldsymbol{m_2}, \boldsymbol{\chi_1}, \boldsymbol{\chi_2}, \boldsymbol{\lambda_1}, \boldsymbol{\lambda_2}$ & $3.94\times 10^6$ & 29.21 & $8.63\times 10^5$ & $2.96\times 10^4$ & TaylorF2 \\
    \bottomrule
    \end{tabular}
\end{table*}

The cost of a search linearly scales with the size of the template bank. We conduct an analysis to test how the time to generate different banks compares to the time for completing the bank's respective search. To explain the computational costs associated for searching different astrophysical populations, we conducted an analysis on six distinct banks, 4 characterized by different parameter (mass, spin, eccentricity, and deformability) generated using our method, one bank using a non-spinning geometric method \citep{2012PhRvD..86h4017B} to compare how a geometric bank search will perform, and one mass only bank generated using an existing stochastic method \citep{2014PhRvD..89h4041A,PhysRevD.93.124007}. All banks were standardized with fixed equal component masses spanning [1,10] $M_\odot$. Additionally, we maintained a consistent minimal match threshold of 0.95 across all banks and a tolerance of 0.01 for the stochastic banks. The lower frequency for the templates was uniformly set to 20Hz, and the upper frequency was 1000Hz for each bank, and parameters such as $\tau_0$ start, crawl, and end remained constant across the stochastic banks. For the other stochastic method, we choose the convergence criteria to be 1000 since that was the most comparable to our banks with a tolerance of 0.01 (showcased in Figure \ref{fig:ff}). For aligned spins, each of the component spins was constrained to [-0.5, 0.5]. For banks accounting for orbital eccentricity ($e$), $e$ was set to vary from [0, 0.2] for a reference frequency of 20 Hz, using the same spin range defined above. For the deformable bank, we set the tidal deformability $\lambda$ to be [0,5000], also using the same aligned spin parameters. Except for the three banks only considering mass, we parallelized three of the banks across six distinct $\tau_0$ bins to speed up the bank generation in wall clock time.

As we increase the number of parameters, we see the banks take a longer time to generate and the number of templates increase by roughly $\mathcal{O}(10)$ for every additional pair of parameter incorporated. Explicitly looking at the TaylorF2 waveform banks, mass only had $\mathcal{O}(10^4)$ and adding a pair of spin parameters drove the bank size up to $\mathcal{O}(10^5)$ and the tidal deformability banks was of size $\mathcal{O}(10^6)$. The eccentric bank and the tidal deformability bank took the longest to generate and had the most amount of templates.

We find that our method performs similarly to existing stochastic and geometric methods in generating a two-parameter mass-only bank. The geometric method produced the bank in the shortest time and the second largest bank. The existing stochastic method produced the most templates, but also took the longest to generate a bank. However, the comparable cost to complete the search was the lowest for sbank. Overall, our method performs just as well as existing methods. The main advantage of our method is its ability to efficiently construct high-dimensional parameter spaces ($N > 4$), in which these other methods are unable to do so. Utilizing the inner product inequality significantly enhances efficiency, particularly for larger, higher-dimensional parameter spaces. To quantify the impact of this efficiency, we estimate the proportion of matches excluded by the inner product inequality. For instance, in one of the largest tidal deformable sub-banks, we observed a $\sim40\times$ reduction in the number of required matches.

Given a fiducial search analyzing a year of data for a three-detector configuration, with each core capable of processing 5000 templates in real-time, we estimate and compare the time required for generating the template bank and running the search \citep{2018PhRvD..98b4050N}. We present these results of the CPU time scaling for all banks in table \ref{tab:scale}. 

We estimate the computational cost for running a non-spinning search is $\mathcal{O}(10^5)$ times greater than the bank generation costs, indicating that the speed of a template bank is negligible. We also find that the geometric bank had more templates and generated 2.3 times faster than our stochastic method. However, the search over time ratio is 2.58 times higher for geometric methods than our stochastic method depicting that faster algorithms are not beneficial towards running the search and the excess number of templates negatively effects the speed of the search. The eccentric and deformable template banks took the longest to complete at around 55 core days. In comparison to the search, the fraction of search time over bank time is $\mathcal{O}(10^4)$ and $\mathcal{O}(10^3)$ less despite having more templates. Since these banks are parallelizable, the cost of generating a bank can always be reduced in wall clock time. Overall, we find that template bank generation is already a negligible computational component of a search, even for high dimensional spaces with millions of templates. Currently, efforts to generate fast template placement algorithms are less necessary than developing more efficient search methodologies, especially for multi-parameter searches.

\section{Examples}\label{sec:ex}

In this section, we present various applications of our stochastic method in different astrophysical searches. The stochastic method has been instrumental in constructing template banks for several types of gravitational-wave searches by utilizing the varying freedom of choice in the parameter space. Below, we discuss specific examples where this method has been successfully implemented.

\subsection{Primordial Black Hole and Sub-Solar Mass Binary Neutron Star}
Detecting low-mass black holes will verify the existence of Primordial Black Holes (PBHs) and provide insights on dark matter distributions \citep{2024NuPhB100316494G,2016MPLA...3150064F,2024arXiv240219468M}. Due to their theorized dynamical formation, searching for these systems require additional parameters to account for the eccentric orbits of PBHs \citep{2024arXiv240107615D}. \citet{PhysRevLett.127.151101} performed a search for PBH systems using the stochastic method outlined in this paper to generate an eccentric template bank. The parameter space included eccentricity $e$ up [0,0.3] with a reference frequency of 10 Hz, primary masses ranged from [0.1,7] $M_\odot$, secondary masses ranged from [0.1,1] $M_\odot$ and used TaylorF2 \citep{Sathyaprakash:1991mt,Droz:1999qx,Blanchet:2002av,Faye:2012we} and TaylorF2e \citep{2019CQGra..36r5003M,2020CQGra..37v5015M,2018CQGra..35w5006M}. To account for the long signal duration, our method allows the option to fix the duration of waveform models by varying the lower frequency. \citet{PhysRevLett.127.151101} used this option to set the wavelength to 512 seconds. Overall, the bank consisted of $7.8 \times 10^6$ templates, where half of the templates included orbital eccentricity.

\begin{figure} 
    \centering
    \includegraphics[height = 7.5cm, width =9cm]{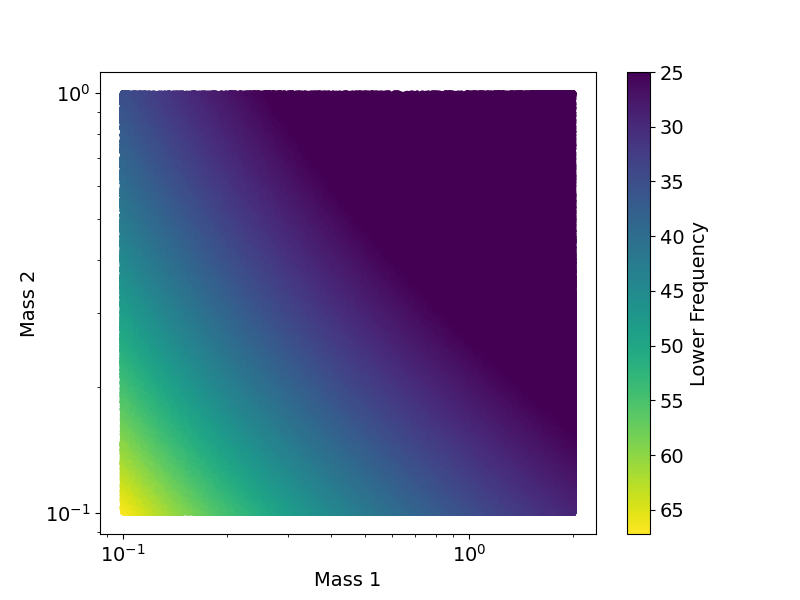} 
    \caption{Sub-solar mass neutron star search template bank with 10 million templates. Mass ranges from 0.1 to 2 for the primary mass and 0.1 to 1 for the secondary mass. The signals are fixed to to 512 seconds by varying the lower frequency cutoff indicated in the color bar.}
    \label{fig:submass}
\end{figure}

Searching for low-mass neutron stars requires a similar template bank as the PBHs search. However, these systems requires consideration for the tidal deformability parameter instead of the eccentricity, as the eccentricity in these systems is expected to be negligible. Discovering neutron stars in the sub-solar mass range could challenge our understanding of their formation or potentially reveal a new class of stars \citep{2022NatAs...6.1444D}. Previous searches neglected the tidal deformability parameter $\lambda$, which lost up to 78.4\% of the total signals \citep{2023PhRvD.107j3012B}. This bank was constructed with tidal deformability ranging from [0,10000] for both ${\lambda_1}$ and ${\lambda_2}$ to account for the loss in SNR, primary mass ranging [0.1,2] ${M_\odot}$, secondary mass ranging  [0.1,1] ${M_\odot}$, both aligned spin $\chi_{1z}$ and $\chi_{2z}$ ranged from [-0.05,0.05], and the approximant was chosen to be TaylorF2 \citep{Sathyaprakash:1991mt,Droz:1999qx,Blanchet:2002av,Faye:2012we}. Similarly to the PBH search, this bank had the template waveforms fixed to 512 seconds to speed up the search. Figure \ref{fig:submass} illustrates the varying lower frequencies of the waveforms used to maintain this duration. Overall, the bank consisted of $1.01 \times 10^7$ templates. \\

\subsection{Eccentric NSBH and BNS Search}

Measuring the orbital eccentricity of a binary provides insights into its formation history \citep{Belczynski:2001uc,Rodriguez:2017pec, Romero-Shaw:2019itr,Sedda:2020wzl,Trani:2021tan,Zevin:2021rtf,Dhurkunde:2023qoe}. Searches for rare binaries with non-negligible eccentricity require additional parameters in the template bank: requiring up to 100x more templates than the typical aligned-spin, quasi-circular searches \citep{Nitz:2021vqh, Nitz:2021mzz,Dhurkunde:2023qoe}. The first modeled search for eccentric spinning neutron-star binaries (BNS and NSBH) using publicly available data from the third observing run of the Advanced LIGO and Virgo observatories, was performed by \citet{Dhurkunde:2023qoe}. \citet{Dhurkunde:2023qoe} utilized the flexible bank generation method described in this work to incorporate eccentricity in the template bank. The template bank comprised of six parameters: component masses ($m_1, m_2$), aligned component spins $(s_{1z}, s_{2z})$ and orbital eccentricity $e_{20}$ (defined at 20 Hz), and an additional angle to account for the orientation of the elliptical orbit w.r.t an observer. Signals within the targeted search region are reliably searched using the inspiral-only TaylorF2Ecc model \cite{Moore:2016qxz}, as the merger phase falls outside the sensitive range of current detectors. The template bank for the search consisted of approximately 6 million templates, and its generation took about a week to complete using 20 cores.

\subsection{Binary Neutron Star Confusion Noise Cleaning}

In the next decade, next-generation ground-based detectors, such as the Einstein Telescope (ET) \citep{2010CQGra..27a5003H,2010CQGra..27s4002P} and Cosmic Explorer (CE) \citep{2019BAAS...51g..35R}, will be available. Those detectors will not just be one order of magnitude more sensitive than the current Advanced LIGO and Virgo detectors, but also can reach 2 or 3 Hz, making the binary neutron stars' signals last several hours or even several days. According to GWTC-3 population results \citep{2023PhRvX..13a1048A}, \cite{2023PhRvD.107f3022W} simulates the mock data for those next-generation ground-based detectors and finds that those BNS signals will form the foreground noise to reduce the detection range of detectors if not removed. This BNS confusion noise will affect Cosmic Explorer most by raising the power spectrum density around 10 Hz and introducing a noise correlation in the detector network. \cite{2023PhRvD.107f3022W} demonstrates a foreground cleaning method using the stochastic template bank for BNS. For bank generation, they choose $[2.4, 60]$ $M_{\odot}$ as the detector-frame total mass range and $[1, 1.636]$ as the mass ratio range according to the BNS population model. With a mismatch of 0.97 using the design sensitivities of ET and CE, the foreground noise-cleaning banks contain $9.57 \times 10^4$ and $1.53 \times 10^5$ templates for CE and ET, respectively. Using these banks to detect and subtract the BNS confusion noise for each detector, they can suppress the total noise to almost the instrument noise level, allowing for near-optimal searches at the following stages. The computational cost of generating and searching with these template banks is lower compared to the full search \citep{2021PhRvD.104f3011L}.

\subsection{LISA Supermassive Black Hole Binaries Search and Inference}

There are ongoing efforts to extend PyCBC to do LISA data analysis, \cite{2024CQGra..41b5006W} demonstrates how to use the stochastic template bank method described in this paper to find supermassive black hole binaries (SMBHBs) and use the corresponding high SNR template as the reference signal in the following heterodyning parameter estimation. Previously, people thought the template-based analysis was not viable for LISA data analysis due to the huge parameter space \citep{2008CQGra..25r4027H}, but \cite{2024CQGra..41b5006W} demonstrates we can still use a sparse template bank to find all SMBHB signals in the LISA mock dataset Sangria. Different from ground-based detectors, LISA waveforms also need to take the orbital motion of spacecraft and time-delay interferometry (TDI) into account, so these make LISA waveforms much more complex. \cite{2024CQGra..41b5006W} uses the BBHx \citep{2020PhRvD.102b3033K} package to generate the LISA-TDI version of the IMRPhenomD waveform and chooses an eight-dimensional parameter space (detector-frame total mass, mass ratio, two aligned spins, ecliptic latitude, ecliptic longitude, polarization angle, and inclination angle) that covers the parameters of SMBHBs in the Sangria dataset. Due to the high SNR of those SMBHB signals and almost equal sensitivities of TDI-A and TDI-E channels, they just generate the TDI-A channel’s template bank and use a mismatch threshold of 0.9 to get the SMBHB template bank consisting of around 50 templates. Finally, they successfully detected all the SMBHB signals in the Sangria dataset.

\subsection{Open Gravitational-Wave Catalog Search for Compact-Binary Mergers}

The fourth open gravitational-wave catalog (OGC) contains the observation of nearly 100 compact binary mergers~\citep{2023ApJ...946...59N}. The search covers a parameter space from neutron star binaries through heavy binary black holes (up to $\sim 1000$ $M_{\odot}$). The OGC search search splits the analysis into four sub-regions, covering BNS, NSBH, BBH, and a focused BBH region where the largest numbers of signals are observed. The analysis based on the open source PyCBC toolkit~\citep{2016CQGra..33u5004U,Davies:2020tsx} and makes use of the flexible template bank generation introduced in this work. This enables the use of multiple waveform approximants depending on the suitability for different parts of parameter space, e.g. the use of TaylorF2~\citep{Sathyaprakash:1991mt,Droz:1999qx,Blanchet:2002av,Faye:2012we} for neutron star binaries, and IMRPhenomD~\citep{Husa:2015iqa,Khan:2015jqa} and SEOBNRv4 for BBH signals~\citep{Taracchini:2013,Bohe:2016gbl}. The stochastic placement algorithm allows for the iterative generation of a larger template bank by adding to a pre-existing template bank. This allows for regions of parameter space that adapt to different requirements, e.g. a higher minimal match for the focused BBH region of 0.995 or the inclusion of tidal deformability for binary neutron star templates.

\subsection{Binaries Contain Neutron Star as Gamma-Ray Burst Progenitors}
Mergers of binary neutron star and neutron-star--black-hole systems have long been suspected to be the production sites of short duration gamma-ray bursts \citep{1984SvAL...10..177B, Eichler1989, Paczynski1986, Paczynski:1991aq, Narayan1992, Nakar:2007yr, Berger2014}. In the case of binary neutron stars, this hypothesis was confirmed by the simultaneous detection of GW event GW170817 and its gamma-ray burst counterpart GRB 170817A \citep{LIGOScientific:2017zic}. The main mechanisms invoked to launch the relativistic jet responsible for the gamma-ray emission involve neutrino pair annihilation or the presence of strong magnetic fields.  Both scenarios require a remnant constituted by a central compact object accreting from a dense torus of matter surrounding it that develops immediately after the merger event.  In the case of two binary neutron stars, the formation of the remnant torus is fueled by the collision of the two neutron stars, while for neutron-star--black-hole mergers the presence of the torus depends on the parameters of the system, \citep[see, e.g.,][for examples of numerical simulations]{Gonzalez:2022mgo, Kyutoku:2021icp}.

When constructing template banks for searches that aim at uncovering GW signals compatible with the time and sky location of gamma-ray bursts \citep{Harry:2010fr, Williamson:2014wma}, the goal is to include all binary neutron star mergers and only those neutron-star--black-hole mergers that result in the formation of an accretion torus, as suggested in \citet{Pannarale:2014rea}.  These are referred to as ``EM-bright'' template banks.  To discriminate between neutron-star--black-hole systems that produce matter surrounding the central remnant black hole and ones that do not, the PyCBC toolkit implements a formula that predicts the remnant mass left behind in the post-merger.  This formula was obtained in \citet{Foucart2018} by fitting results of numerical-relativity neutron-star--black-hole merger simulations; it returns the remnant mass $M_{\textnormal{rem}}$ given the following parameters of the binary system: its symmetric mass ratio, the radius of the black hole's innermost stable circular orbit, and the neutron star compactness \citep[see Eq.\,(7) in][for details]{Foucart2018}.  These banks are therefore built by prescribing priors for masses and spins of neutron stars and black holes, and applying the constraint that the remnant mass surrounding the central black hole is non-vanishing in the case of neutron-star--black-hole systems.  Given a draw from the priors, there are two operations in this process that depend on the neutron star equation of state.  1) Deciding whether or not each compact object is a neutron star, so that binary neutron stars are added to the bank if necessary, and binary black holes are discarded.  2) In the case of a neutron-star--black-hole binary draw, determining the neutron star compactness corresponding to its mass in order to apply the constraint $M_{\textnormal{rem}}>0$.  Building an EM-bright bank therefore also requires an additional input in the form of a table with mass and compactness values that represent a non-rotating\footnote{Recall that the $M_{\textnormal{rem}}$ formula does not depend on the neutron star spin.} neutron star equilibrium configuration; this needs to be built externally by the user, see, e.g., \url{https://compose.obspm.fr/}.  Finally, in the case of black hole spins that are not aligned to the orbital angular momentum, we replace the radius of the innermost stable circular orbit with its tilted analogue, the radius of the innermost stable spherical orbit, as detailed in Appendix A of \citet{Stone:2012tr}.

Two neutron-star--black-hole banks were produced: one with with the EM-bright constraint and one without it. The design parameters were the same for both banks, with the exception of the remnant mass, which was fixed to be strictly greater than zero for the EM-bright bank. In this case, the 2H piecewise polytropic equation of state \citep{Read:2009yp} was adopted.  This choice was driven by the fact that this equation of state sets a high maximum neutron star mass ($\sim 2.83\,M_\odot$) and it favours tidal disruption because of the high neutron star compactess values it yields, compared to other equations of state.  In this sense it is a conservative choice, that is, it makes the EM-bright constraint as loose as possible.  The black hole mass and spin spanned from $2.83\,M_\odot$ to $25\,M_\odot$ and from 0.0 to 0.98, respectively. The neutron star mass ranged from $1.0\,M_\odot$ to $2.83\,M_\odot$, while the aligned spin ranged was 0.0 to 0.05. All priors in these intervals were taken to be uniform.  Additionally, we used a minimal match of 0.97 and a lower frequency of $27.0$ Hz, and we set the PSD to be the theoretical Advanced LIGO O4 sensitivity \citep{Abbott2020}. The full bank consisted of 193,235 templates, while the EM-bright bank had 143,530 templates, resulting in the exclusion of 49,705 neutron-star--black-hole templates representing the portion of parameter space where merger events do not produce a torus as a remnant.

\section{Conclusion} \label{sec:conclusion}
We introduce an efficient and flexible method for stochastic template bank generation that produces self-consistent banks for varying tolerance and minimal match conditions. Our method has already been used for various searches for GWs from different types of compact binary sources such as eccentric BNSs and NSBHs \citep{2023arXiv231100242D}, sub-solar mass primordial black holes \citep{PhysRevLett.127.151101}, low-mass BNSs, studies of next-generation ground-based detectors' BNS signal detection problem \citep{2023PhRvD.107f3022W,2021PhRvD.104f3011L}, LISA supermassive BBHs \citep{2024CQGra..41b5006W}, and the fourth open gravitational-wave catalog \citep{2023ApJ...946...59N}. We find that existing methods are robust and versatile enough for various unique astrophysical scenarios.

We have demonstrated how the number of parameters in a search scales the computational cost. Compared to a non-spinning bank with only two parameters (masses only), we find an increase of up to $\mathcal{O}(10^2)$ templates as additional parameters are included in the bank. For future observatories, aligned-spin banks will get bigger by up to two orders of magnitude compared to the current banks \citep{Dhurkunde:2022aek}. Fast bank generation algorithms alone will not inherently lead to faster searches. The most pressing need for improvement is developing faster or optimizing existing search methods. Ongoing efforts to utilize various hierarchical methods could make current searches up to 20 times faster \citep{Dhurkunde:2021csz, Soni:2021vls}. If efforts to procure faster searches succeed, the cost of template banks will once again become considerable, and efforts to create faster methods will become critical. 

In conclusion, while our method demonstrates significant improvements in template bank generation, it is important to consider the limitations and potential optimizations for utilizing our method for future optimized searches. Stochastic methods can be limited when the waveform approximant takes a long time to generate, as the time to generate a waveform linearly scales with the template bank. If the waveform generation time is comparable to the search time, geometric methods might be more appropriate. However, if a search requires a non-trivial parameter space and utilizes slow waveform approximants, optimizing stochastic bank methods becomes necessary. Our method is currently written for CPU use. This method could potentially be optimized by parallelizing the template acceptance proposals on GPUs to help speed up the generation. This optimization might help offset the additional time required for slow waveform approximants.

The bank generation code \texttt{pycbc\_brute\_bank}, which uses the method described in this paper, is at \url{https://github.com/gwastro/pycbc/blob/master/bin/bank/pycbc_brute_bank}.

\section*{Acknowledgments}
KK and AHN acknowledge support from NSF grant PHY-2309240. KK and AHN acknowledge the support from Syracuse University for providing the computational resources through the OrangeGrid High Throughput Computing (HTC) cluster. Additionally, KK expresses gratitude to Ananya Bandopadhyay and Aleyna Akyüz for their valuable feedback and insightful comments, which greatly improved the quality of this paper. RD and SW acknowledges Max Planck Gesellschaft for support. MC acknowledges the support through the grant PID2021-127495NB-I00 funded by MCIN/AEI/10.13039/501100011033 and by the European Union, and the Astrophysics and High Energy Physics programme of the Generalitat Valenciana ASFAE/2022/026 funded by MCIN and the European Union NextGenerationEU (PRTR-C17.I1).  FP acknowledges support from the ICSC - Centro Nazionale di Ricerca in High Performance Computing, Big Data and Quantum Computing, funded by the European Union - NextGenerationEU and support from the Italian Ministry of University and Research (MUR) Progetti di ricerca di Rilevante Interesse Nazionale (PRIN) Bando 2022 - grant 20228TLHPE - CUP I53D23000630006.

\newpage

\bibliography{sample631}{}
\bibliographystyle{aasjournal}

\end{document}